\documentclass[12pt]{iopart}
\usepackage{graphicx}
\begin{document}
\title[A solvable problem in statistical mechanics]{A solvable problem in statistical mechanics: the dipole-type Hamiltonian mean field model}
\author{Boris Atenas and Sergio Curilef}
\address{Departamento de F\'\i sica, Universidad Cat\'olica del Norte, Avenida Angamos 0610, Antofagasta, Chile}
\ead{scurilef@ucn.cl}
\vspace{10pt}
\begin{indented}
\item[]December 2017
\end{indented}

\begin{abstract}
The present study regards the zeroth order mean field approximation of a dipole-type interaction model, which is analytically solved in the canonical and microcanonical ensembles. After writing the canonical partition function, the free and internal energies, magnetization and the specific heat are derived and graphically represented.
A crucial derivation is the calculation of the free energy, which is variationally evaluated, and it is shown that the exact result coincides with the approximate trend when $N$ tends to infinity. In the microcanonical ensemble,
the entropy as other thermodynamic properties are calculated. We
notice that both schemes coincide in equilibrium.
\end{abstract}

\section{Introduction}
\label{intro}
In statistical mechanics, systems with analytical solutions are not very common; even more, if we take into account systems with long-range interactions, solutions become most difficult to obtain. However, the hard sphere, Van der Waals, Weiss, Bragg-Williams, Bethe, Landau, Ising, HMF are well established models and their analytical solutions are known~\cite{reichl,soto,Glasser}.
In the field of applications, the Ising model represents the most relevant tool for studying magnetic properties and the statistical behavior of many-body systems in the most  wide context\cite{DS,Tatekawa,cdr,Pluchino, Curilef}. 

In this paper,  we make an effort to discuss another possible variation of the Ising model that includes long-range interactions, which seems to be interesting from the atomic scale to the astronomical scale.  A typical
consequence, of including long-range interacting particles in systems, is the loss of standard thermodynamic properties such as loss of additivity and/or loss of extensivity, etc. The loss of additivity occurs when they cannot be trivially separated into independent subsystems, which is efigxplained by the existence of underlying interactions or correlation effects whose characteristic lengths are comparable or larger than the system linear size\cite{TDauxois,Levin,delPinoPRB2007}. The loss of additivity is frequently accompanied by the loss of extensivity. A thermodynamic variable, such the energy or the entropy, is extensive providing that the variable is proportional to the number of elements and the intensive variables are kept constant. To illustrate these concepts, let us define a type of variation of Ising model with vanishing magnetization, $M=\sum_i s_i = 0$ that involves long-range interactions given by 
\begin{equation}
H = \lambda\sum_{i=1}^{N-1}\sum_{j=i+1}^N \frac{{s}_i{s}_j}{|i-j|^\alpha}
\end{equation}
where $\lambda$ is a constant, $s_i = \pm 1$, $\forall i$ and $|i-j|$ represents the distance between two sites. 
 Now, let us divide the system into two subsystems, I and II, each is composed of $N/2$ sites, where all spins in subsystem I are up;
while those in subsystem II are down. Total energy of the system is $E = 0$. Nevertheless, the energies of the two subsystems, $E_I=\lambda\sum_{i=1}^{N/2-1}\sum_{j=i+1}^{N/2} s_is_j/|i-j|^\alpha$ and $E_{II}=\lambda\sum_{i=N/2}^{N-1} \sum_{j=i+1}^{N}s_is_j/|i-j|^\alpha,$ satisfy $E_I=E_{II}\neq 0$. Since the sum of the two energies $E_I+E_{II}$ is not equal to the total energy $E$, therefore the system is clearly nonadditive. In general, $E < E_I + E_{II}$.
Despite the persistence of the loss of additivity, with the aid of a Kac-like scaling\cite{kac} is possible to recover extensivity\cite{reichl,delPinoPRB2007,kac,Atenas2017}. This kind of scaling presents a standard thermodynamic structure because it preserves the Euler and Gibbs-Duhem relations recovering\cite{Curilef,delPinoPRB2007}, for instance the linearity of the thermodynamic properties of a system with long-range interactions by a scaling Hamiltonian\cite{Pluchino,delPinoPRB2007,kac,Atenas2017}.
In addition, several particular properties related to nonequilibrium behavior of systems with interacting particles,
as the relaxation to thermal equiibrium in $N$-particle Hamiltonian, have been recently the subject of an intense debate\cite{TDauxois}. 
The relaxation time has been shown to be long and to increase with the number of particles. The system evolves very slowly, on time scales diverging
with $N$, towards Boltzmann-Gibbs equilibrium. States that evolve on time scales that diverge with $N$ are called ``quasi-stationary''. Some of them were previously identified and characterized \cite{Atenas2017} for the dipole-type Hamiltonian mean field (d-HMF) model. Such a discussion is pertinent due to several physical applications\cite{TDauxois}.

Nevertheless, in the present paper we are not interesting in the discussion of nonequilibrium, the main aim is to discuss the behavior and properties of the system that can be derived from the statistical ensembles in equilibrium.
It is generally accepted that two phases appear in the phase diagram, which depend on the energy and the temperature. At low energy, a phase identified by the presence of a single cluster of particles arises by floating in a diluted homogeneous background. At high energy a homogeneous phase is recovered; the cluster disappears and the particles move (almost) freely. For a pertinent transition region below a critical value, the system is characterized by the microcanonical ensemble with negative specific heat and the resulting instability is extremely relevant\cite{Levin} because of its strong implications on both experimental and theoretical features.

We organize the paper as follows. In section~\ref{model}, we introduce the zeroth order approximation from the potential energy for classical electric dipoles, and we build the d-HMF model proposed previously\cite{Atenas2017}. In section~\ref{results}, we calculate the canonical partition function of the system using approximate and variational methods to derive explicitly thermodynamics quantities, as the free energy, magnetization, internal energy and specific heat.
In section~\ref{microcanonical}, we use the microcanonical ensemble to compute the entropy and show that the derivation of the thermodynamics  coincides with the one derived in the previous section.
Finally, in section~\ref{concluding}, we make a summary of the results and draw some conclusions.

\section{The Model}\label{model}

\begin{figure}[b!]
\centering
\includegraphics[scale=0.6]{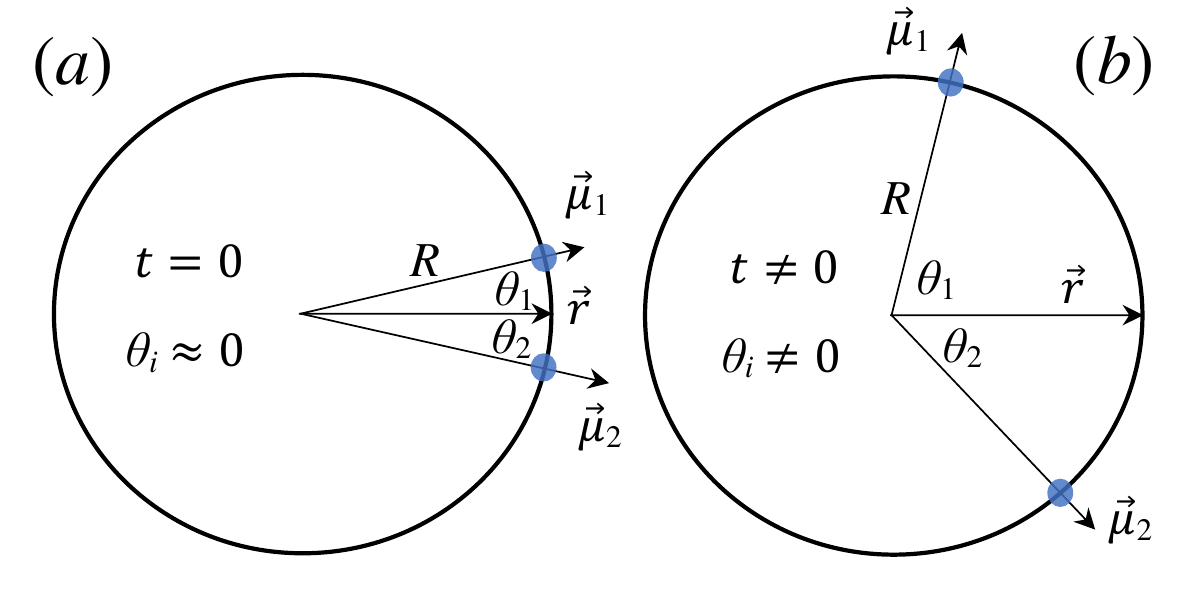}
\caption{The figure shows the initial conditions and a possible evolution for a couple of dipoles in a ring.  ({\emph{a}}) Initial orientations for $t=0$  for the configuration of two dipoles. ({\emph{b}}) Possible configuration for $t\neq0$.}\label{ring}
\end{figure}

If we consider a system of two interacting dipoles $i$ and $j$ with momenta $\mu_i$ and $\mu_j$ respectively, the potential energy is given by
\begin{equation}
U= - \frac{1}{4\pi\epsilon_0}\frac{3(\vec{\mu}_i\cdot \hat{r})(\vec{\mu}_j\cdot \hat{r})-\vec{\mu}_i\cdot\vec{\mu}_j}{\vert\vec{r}_i-\vec{r}_j\vert^3},
\end{equation}
where $\hat{r}$ is a unit vector and the vector  of an arbitrary direction and $\vec{r}_i$  ($\vec{r}_j$) corresponds to the position  of the particle $i$ ($j$).
Besides, we define the moment of the electric dipole $i$ as  $\vec{\mu}_i = q \vec{a}$, where $q$ is the modulus of each charge of the dipole and the modulus of vector $\vec{a}$ represents the separation of charges  of the dipole;  and the vector character of  $\vec{a}$ emphasizes its orientation. Therefore,
$\vec{\mu}_i \cdot \hat{r} = \mu \cos {\theta_i}$ and $\vec{\mu}_i \cdot \vec{\mu}_j = \mu^2 \cos(\theta_i-\theta_j)$.
In addition, we remark that $\mu=|\vec{\mu}_i|$, the vectors $\vec{r}_i$ and $\vec{\mu}_i$ are parallel between them for all $i$.
We illustrate the model in Figure \ref{ring} for two dipoles in a ring. The system is formed by $N$ dipoles that can be distributed in a ring to have a perspective of the total system.
On one hand, the distance between dipoles in a ring, with radius $R$, can be written as
\begin{eqnarray}
r&=&\vert\vec{r}_i-\vec{r}_j\vert \\
&=&\vert R\cos\theta_i \hat{i}+R\sin\theta_i \hat{j}-R\cos\theta_j \hat{i}-R\sin\theta_j \hat{j} \vert \\
&\approx&\sqrt{2}R( 1-\cos\theta_i \cos\theta_j -\sin\theta_i \sin\theta_j +\delta)^{1/2}.
\end{eqnarray}
The parameter $\delta$  corresponds to a softening parameter that is usually introduced\cite{Tatekawa}  to avoid the divergence of the potential at short distances. {In addition, we consider the identity}
\begin{equation}
\cos(\theta_i-\theta_j)=\cos\theta_i \cos\theta_j+\sin\theta_i \sin\theta_j,
\end{equation}
which leads us
\begin{eqnarray}
r^{-3}&\approx & (2R^2)^{-3/2}(1-\cos(\theta_i-\theta_j)+\delta)^{-3/2} \nonumber\\
&\approx & (2R^2\delta)^{-3/2}\left(1-\frac{\cos(\theta_i-\theta_j)}{\delta}+\frac{1}{\delta}\right)^{-3/2}.
\end{eqnarray}
By mean a binomial expansion in $\frac{1}{\delta}$, the interaction potential energy between dipoles can be written as
\begin{equation}
U\!\!\approx\! \! \frac{\!-\mu^2}{4\pi\epsilon_0(2R\delta)^{3/2}}\!\!\left(\!\frac{}{}\!3\cos\!\theta_i \!\cos\!\theta_j\! -\!\cos(\!\theta_i\!\! -\!\!\theta_j)\right)\!\! \left(1\!\!-\!\!\frac{3}{2} \frac{\cos(\theta_i\!-\!\theta_j)}{\delta}\!+\!O(\!\delta ^{-2}\!)\!\right)\!.
\end{equation}
Taking the large $\delta$ limit, in the zeroth order approximation, the potential energy of $N$ dipoles becomes
\begin{equation}
U\!\approx \! \frac{\lambda}{2N} \displaystyle \!\sum_{i\neq j}^{N}\!\left(\!\frac{}{}\!\!\cos(\theta_i \!-\!\theta_j)\!\!-\!\!3\cos\theta_i\! \!\cos\theta_j\!\!\right),
\end{equation}
where $\lambda$ is the coupling constant, whose sign depends on the initial orientation of the dipoles. If the coupling, $\lambda$, is positive, then the system is ferromagnetic; however, if it is negative, the system is anti-ferromagnetic. The constant $N$ in the denominator stands for a scaling to guarantee properly the extensivity of the system \cite{reichl,Curilef,delPinoPRB2007,kac,Atenas2017}.

For theoretical modeling of systems with long range interactions, we take a system of $N$ identical coupled particles, dipole type, with a mass equal to $1$,  whose dynamics evolves in a periodic cell described by a 1D d-HMF model given by\cite{Atenas2017}
\begin{equation}\label{powerlaw}
 H \!= \!\!\sum_{i=1}^{N}\!\frac{p_i^2}{2}\!+\! \frac{\lambda}{2N} \!\sum_{i\neq j}^N \left(\frac{}{}\!\!\cos(\theta_i\!-\!\theta_j)\!-\!3\cos\theta_i \cos\theta_j\!-\!\Delta_{i,j}\right),
\end{equation}
where the variable $p_i$ represents the momentum of the particle $i$ and $\theta_i$ represents its corresponding angle of orientation. With the integer $i\in [1,N]$,  $N$ is the size of the system. The parameters $\lambda$ and  $\Delta_{i,j}$ denote the coupling and initial conditions, respectively. The parameter $\Delta_{i,j}$ suitably establishes the zero of the potential energy as
\begin{equation}
\Delta_{i,j}=\cos(\theta_{0i}\!-\!\theta_{0j})\!-\!3\cos\theta_{0i} \cos\theta_{0j},
\end{equation}
where  the set of angles $\{\theta_{0k}\}$ denotes the initial orientations of the particles. Due to the nature of the model, it has not been taken into account any dependence on the  distance, just the dependence on the orientation is considered in the dipole description.

\section{Canonical ensemble}\label{results}
\subsection{Equations of Motion}
The interacting part of the model is commonly expressed in terms of the spin vector related to each particle and it is given by
$\overrightarrow{m}_i=(\cos\theta_i,\sin\theta_i).$
Therefore, we can introduce the total spin vector
\begin{eqnarray}\label{totalspin}
\overrightarrow{M} &=& \frac{1}{N}\sum_{i=1}^N\overrightarrow{m}_i \label{defM} \\
&=&(M_x,M_y) \label{defMxy} \\
&=& m\exp(i\phi),
\end{eqnarray}
where $(M_x,M_y)$ and $m$ are the components and the modulus of the vector $\overrightarrow{M}$, respectively, and $\phi$ denotes the phase of the order parameter. The equation of motion is
\begin{equation}\label{eqmot}
\dot{p}_i=-\lambda\left(2M_x\sin\theta_i +M_y\cos\theta_i\right)
\end{equation}
and the potential energy can be written as
\begin{equation}
U = - N\frac{\lambda}{2} \left(2M_x^2-M_y^2+\Delta\right), \label{epot}
\end{equation}
where $\Delta =\sum_{i,j}\Delta_{i,j}/N^2$. As aforementioned, this definition is crucial for defining the energy of the system. In consequence, at $t=0$ potential energy is zero and the kinetic energy is maximum and coincides with the value of the total energy, in particular the value $\Delta=-2$ is suitable when initially the dipoles are all parallel, namely $\theta_{0i} =0$.

\subsection{The partition function}
In the canonical ensemble, the partition function is
\begin{eqnarray}
Z(\beta,N)\!&=&\!\!\int \mathrm{d}^{\!N}\!p_i\, \int  \mathrm{d}^{\!N}\!\theta_i e^{-\beta H}\\ 
&=&Z_K(\beta,N)Z_U(\beta,N),
\end{eqnarray}
where $Z_K(\beta,N)$ is the kinetic part of the integral and the $Z_U(\beta,N)$ is the interacting part.
On one hand, the kinetic part is well known and corresponds to
\begin{eqnarray}
Z_K(\beta,N)&=&\!\int \mathrm{d}^N\!p_i  \exp\left(\!-\frac{\beta}{2} \sum_i p_i^2\right)\\
&=&\left(\frac{2\pi}{\beta}\right)^{N/2}.
\end{eqnarray}
On the other hand, the interacting part is
\begin{equation}
{Z_U(\beta,N)=\int\mathrm{d}^N\theta_i \mathrm{exp}\left(\beta N\frac{\lambda}{2}(2M^2_x-M^2_y+\Delta)\right)},
\end{equation}
which we rephrase to remark two Gaussian in the argument of the integral
\begin{equation}
Z_U(\beta,N)= e^{\beta N\frac{\lambda}{2}\Delta} \int\mathrm{d}^N\theta_i e^{\beta N{\lambda}M^2_x} e^{-\beta N\frac{\lambda}{2}M^2_y}.\label{zv}
\end{equation}
 To solve the above integral, let us change the Gaussian terms to equivalent expressions using the integral of the \ref{app1}.
 If we substitute Eqs.(\ref{emxx}) and (\ref{emyy}) in Eq.(\ref{zv}),
 we have
\begin{equation}
Z_U(\beta,N)\!=\!e^{\frac{\Delta}{2}\beta \lambda N}\sqrt{\frac{1}{2\pi^2}}\!\int^{\infty}_{-\infty}\!\!\mathrm{d}y e^{-\frac{y^2}{2\beta \lambda N}}\! \int^{\infty}_{-\infty} \!\! \mathrm{d}x e^{-\beta \lambda N x^2 } \!\int\!\!\mathrm{d}^N\!\theta_i e^{2\beta \! \lambda x\! \sum_i\!\cos\theta_i}\!.
\end{equation}
Therefore, we write
\begin{equation}
Z_U(\beta,N)\!=\!\sqrt{\frac{\beta \lambda N}{\pi}}e^{\frac{\Delta}{2}\beta \lambda N}\!\!\int_{-\infty}^\infty \!\! \!\mathrm{d}x\, e^{-\beta \lambda N x^2} (2\pi\mathrm{I}_0(2\beta \lambda x))^N,
\end{equation}
where $\mathrm{I}_k(y)$ is the modified Bessel function of the k\emph{th}-order.
Then the partition function can be finally written as
\begin{equation}
Z\!=\!\sqrt{\frac{\beta \lambda N}{\pi}}e^{\frac{\Delta}{2}\beta \lambda N}\! \left(\frac{2\pi}{\beta}\right)^{\! \! \frac{N}{2}}\!\! {\cal F}_N(\beta \lambda), \label{function}
\end{equation}
where ${\cal F}_N(\beta \lambda)$ stands for the integral
\begin{equation}
{\cal F}_N(\beta \lambda)= \int_{-\infty}^\infty \!\mathrm{d}x\, e^{-N(\beta \lambda x^2\!-\ln(2\pi\mathrm{I}_0(2\beta \lambda x)))}. \label{intF}
\end{equation}

\subsection{Evaluating the partition function}\label{evalpf}
\begin{figure}
\centering
\includegraphics[scale=0.45]{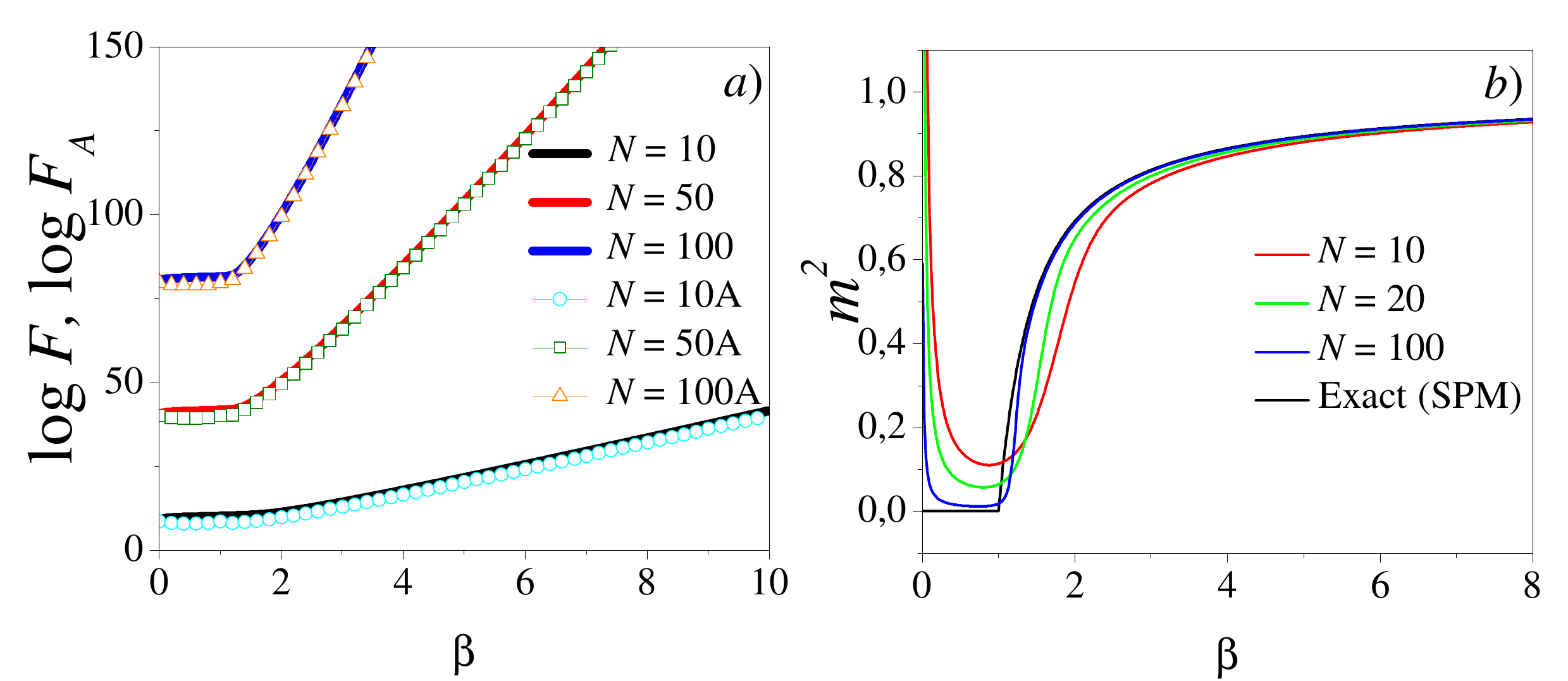}
      \caption{The figure depicts as functions, exact ($F_N(\beta \lambda)$) and approximate as explained, equivalent to the saddle point Method (SPM) are closer as $N$ increases. In ({\emph{a}}) ${\cal F}_N(\beta \lambda)$  is compared to ${\cal F}(N,\beta \lambda)$ for several values of $N$. In ({\emph{b}}), $dX(N,\beta \lambda)/d\beta$ for differents values of $N$ is compared to $dX_{N\rightarrow\infty}(\beta \lambda)/d\beta$ (Exact), which coincide with the square of magnetization as function of inverse temperature $M^2(\beta \lambda)$. }\label{approx}
\end{figure}

If we take the function  $f(x)=N (\beta \lambda x^2-\ln(2\pi\mathrm{I}_0(2\beta \lambda x)))$, we can define an extremum of the function in $x_0=M_x =m=\frac{\mathrm{I}_1(2\beta \lambda x_0)}{\mathrm{I}_0(2\beta \lambda x_0)}$ that  corresponds to the magnetization. The derivative of the second order of the function $f(x)$ evaluated in $x_0$ is given by
\begin{equation}
f''(x_0)=4N\beta \lambda\left(1+\beta \lambda(m^2-1)\right))
\end{equation}
Therefore, the function $f(x)\simeq f(x_0)+\frac{1}{2}(x-x_0)^2f''(x_0)+...$ is explicitly written as
\begin{equation}
f(x)\!\simeq \!N\!\beta\! \lambda x_0^2\!-\!N\!\ln(2\pi\mathrm{I}_0(2\beta \lambda x_0))\!+\!\frac{1}{2}(x\!-\!x_0)^2 4N\!\beta \lambda\left(\!1\!+\!\beta \lambda(\!m^2\!-\!1\!)\!\right)\!,\label{approxf}
\end{equation}
which can be compute by the next approximation
\begin{eqnarray}
\int_{-\infty}^{\infty}\mathrm{d}x \: e^{-f(x)}
&\approx&\int_{-\infty}^{\infty} \mathrm{d}x \: e^{-f(x_0)-\frac{1}{2}(x-x_0)^2f''(x_0)}\\
&\approx&e^{-f(x_0)}\sqrt{\frac{2\pi}{f''(x_0)}}\;\;\;\;=\;\;\;{\cal F}(N,\beta \lambda),
\end{eqnarray}
where   ${\cal F}(N,\beta \lambda)$ corresponds to the approximation of the function ${\cal F}_N(\beta \lambda)$, which is obtained from the evaluation of the integral (\ref{intF})  using the approach given by Eq.(\ref{approxf}). Therefore,
 ${\cal F}_N(\beta \lambda)$ coincides with ${\cal F}(N,\beta \lambda)$ whenever  $N\rightarrow\infty$, the limiting case.
In Figure \ref{approx}({\emph a}), we depict ${\cal F}_N(\beta \lambda)$ compared to ${\cal F}(N,\beta \lambda)$ as a function of $\beta \lambda$  to illustrate how good the approximation is. We see that both values, exact and approximate, are closer as $N$ increases.
The same effect, related to the size of the system, is shown in Figure \ref{approx}({\emph b}) for the defined functions as:
\begin{eqnarray}
X_N(\beta \lambda) &=&\frac{1}{N}\ln \left( {\cal F}_N(\beta \lambda)\right)\\
X(N,\beta \lambda) &=&\frac{1}{N}\ln \left( {\cal F}(N,\beta \lambda)\right).
\end{eqnarray}
As before, $X_N(\beta \lambda)$ and $X(N,\beta \lambda) $ coincide whenever $N\rightarrow\infty$.

Therefore, the partition function can be explicitly written as follows
\begin{equation}
Z\!=\left(\frac{2\pi}{\beta}\right)^{N/2}\!e^{\frac{\Delta}{2}\beta \lambda N}\:\:e^{-N\beta \lambda x_0^2+N\ln(2\pi\mathrm{I}_0(2\beta \lambda x_0))}\frac{1}{\sqrt{2\left(1+\beta \lambda(m_0^2-1)\right)}},
\end{equation}
where $x_0$ is the extremum of the function $f(x)$.
\subsection{Thermodynamics}
Calculating $\ln Z$, we get
\begin{eqnarray}
\ln Z=\frac{N}{2}\ln\!\!\!\!&&\!\!\!\!\left(\frac{2\pi}{\beta}\right)+\frac{1}{2}\ln\left(\frac{\beta \lambda N}{\pi}\right)+\frac{\Delta \beta \lambda N}{2}-N\beta \lambda x_0^2\nonumber\\
&+&N\ln(2\pi\mathrm{I}_0(2\beta \lambda x_0)\!)\!+\!\frac{1}{2}\ln\left(\frac{\pi}{2N\beta \lambda\left(\!1\!+\!\beta \lambda(m^2\!-\!1)\right)}\!\right).
\end{eqnarray}
The limiting case of this quantity per particle is given by
\begin{equation}\label{fi}
\displaystyle\lim_{N\longrightarrow \infty}\frac{\ln Z}{N}=-\frac{1}{2}\ln\left(\frac{2\pi}{\beta}\right)-\frac{\Delta \beta \lambda}{2}+\beta \lambda x_0^2-\ln(2\pi\mathrm{I}_0(2 \beta \lambda x_0)).
\end{equation}
This last expression was evaluated in the thermodynamic limit, $N\rightarrow \infty$\cite{stanley}, at a point of local value given by $x=x_0$.
The free energy per particle $\varphi=-\displaystyle\lim_{N\longrightarrow \infty}\ln Z/N$ is commonly expressed as the extremal problem
\begin{equation}
\varphi(\beta,N)\!=-\!\frac{1}{2}\ln\frac{\beta}{2\pi} \!-\!\frac{\Delta}{2} \lambda\beta 
-\!\inf_{x\geq 0}\![-\!\beta \lambda x^2\!+\!\ln (2\pi\mathrm{I}_0(2\beta \lambda x))].
\end{equation}
As mentioned before, the solution of the extremal is obtained by
\begin{equation}\label{XSolution}
x =\frac{\mathrm{I}_1(2\beta \lambda x)}{\mathrm{I}_0(2\beta \lambda x)}.
\end{equation}
The critical inverse temperature is $\beta_c=1$.

If $\lambda <0$, then the equation has a trivial solution, $x=0$. In contrast, if $\lambda >0$, then the equation has a set of values for $x$ and $\beta$, which defines the solution of the problem. Finally, the internal energy per particle is obtained  as a function of the inverse temperature and magnetization
\begin{equation}\label{ASolution}
\varepsilon=\frac{\partial \varphi(\beta,N)}{\partial \beta} = \frac{1}{2\beta}-\lambda\left( m^2+\frac{\Delta}{2}\right),
\end{equation}
where $m$ is the solution of the extremal problem and corresponds to the solution that we derive from the canonical ensemble. From this time forth, we set $\lambda=1$.
In Figure \ref{approx}
we depict the equilibrium magnetization $m$ as a function of the internal energy $\varepsilon$. The analytical solution is obtained from the canonical ensemble given by Eq.(\ref{ASolution}). The critical point is located at $\varepsilon_c=3/2$, which is twice the value obtained for the HMF model.

\begin{figure}[t]
\centering
 \includegraphics[scale=0.45]{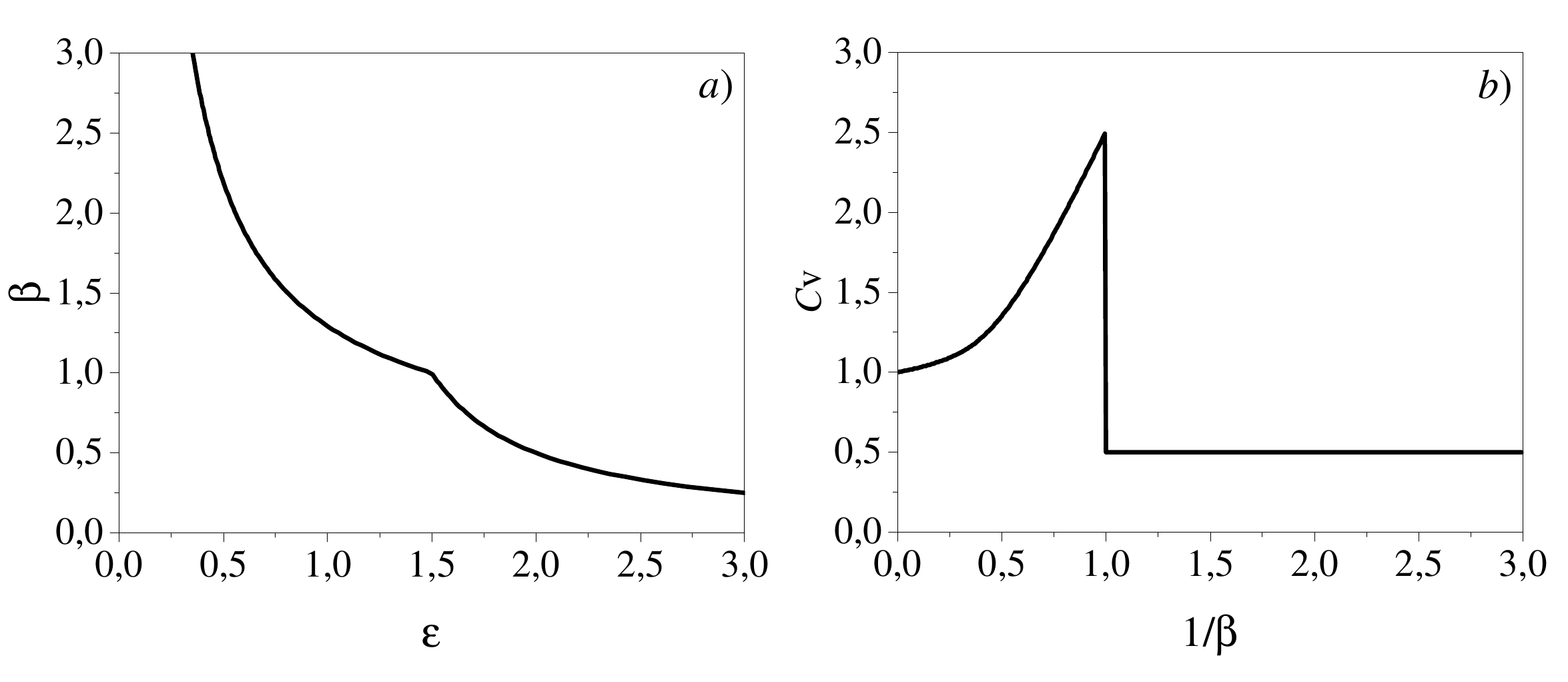}
      \caption{We depict ({\emph{a}})  the internal energy $\varepsilon$ as a function of  $\beta$. The critical point is in $\varepsilon_c = 3/2$, $\beta_c= 1$ for $\Delta=-2$ and $\lambda=1$. Additionally, we depict  ({\emph{b}}) the specific heat $C_v$ as a function of $1/\beta$. It is notorious that in both cases the critical point is located in $\beta_c=1$ }\label{Energy}
\end{figure}

Moreover,  in Figure \ref{Energy} we show ({\emph{a}}) the internal energy $\epsilon$ as a function of $\beta$ and  ({\emph{b}}) the specific heat $C_v$ as a function of $1/\beta$. Once again, it is notorious that  the critical values emphasized in the Figure \ref{Energy}  are: $\beta_c= 1$ and  $\varepsilon_c = 3/2$. We notice that the behavior of thermodynamical functions as $\varepsilon$ and $C_v$ define two different regions. In particular,  Figure \ref{Energy}({\emph{b}}) shows, in the one hand, the specific heat that grows as the temperature (inverse of $\beta$) increases. In other hand, the specific heat keeps constant in $C_v=1/2$, which corresponds to an ideal gas in one dimension.

\section{Microcanonical entropy}\label{microcanonical}

In the microcanonical ensemble, from the Hamiltonian of the Eq.(\ref{powerlaw}), it is possible to iterate the number of microstates by means of
\begin{equation}
\Omega(E,N)=\int\displaystyle\prod_{i=1}^N \mbox{d}p_i\mbox{d}\theta_l\delta(E-H_N(\theta_i,p_i))
\end{equation}
and by introducing the Dirac delta identity on $K$, we have:
\begin{equation}
\Omega(E,N)=\int \mbox{d}K\displaystyle\prod_{i=1}^N \mbox{d}p_i\mbox{d}\theta_i\delta\left(K-\sum_{j=1}^N\frac{p^2_j}{2m}\right)\delta(E-K-U\{\theta_i\}).
\end{equation}
This expression can be separated in the kinetic and configuration parts; therefore, taking $E=K+U$, we have
\begin{eqnarray}
\Omega_{\mathrm{kin}}(K)&=&\int \displaystyle \prod_{i=l}^N \mbox{d}p_i\delta\left(K-\sum_{j=1}^N\frac{p^2_j}{2m}\right),\\
\Omega_{\mathrm{conf}}(E-K)&=&\int \displaystyle\prod_{i=1}^N \mbox{d}\theta_i \delta(E-K-U\{\theta_i\})
\end{eqnarray}
where  $\Omega (E,N)=\int \mbox{d} K \Omega_{\mathrm{kin}}(E,N)\Omega_{\mathrm{conf}}(E-K)$. 
The kinetic integral is well known and given by
\begin{equation}
\Omega_{\mathrm{kin}}(K)=\frac{\pi(2\pi K)^{N/2-1}}{\Gamma(N/2)}.
\end{equation}
Then, using the property $\ln\Gamma(N)=\left(N-\frac{1}{2}\right)\ln N-N+\frac{1}{2}\ln(2\pi)$, and after some manipulation, for large $N$, we obtain
\begin{equation}
\ln\Omega_{\mathrm{kin}}(K)\simeq\frac{N}{2}\left(1+\ln(2\pi)+\ln \frac{2K}{N}\right).
\end{equation}
Defining  $u=2K/N$, $\Omega_{\mathrm{kin}}(K)$ can be expressed as
\begin{equation}
\Omega_{\mathrm{\mathrm{kin}}}(K)\simeq \exp\left(\frac{N}{2}\left(1+\ln(2\pi)+\ln u\right)\right).
\end{equation}

\begin{figure}[t]
\centering
    \includegraphics[scale=0.45]{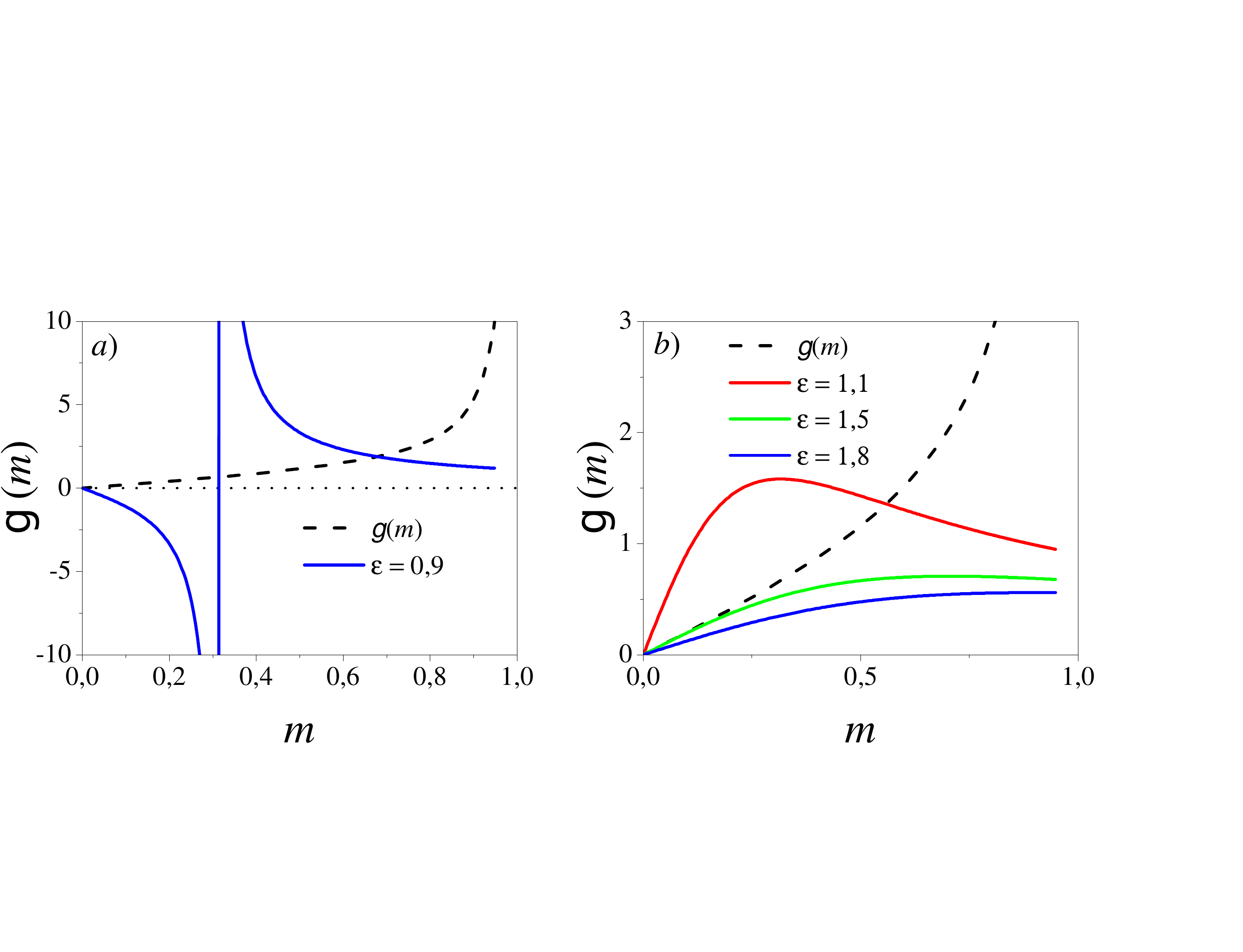}
\caption{We depict $g(m)$, as a function of $m$ for values of ({\emph{a}}) $\varepsilon$ =0.9 and  ({\emph{b}}) $\varepsilon$ =1.1, 1.5, 1.8; $\Delta=-2$ and $\lambda=1$. The value $\varepsilon =1$ changes the trend of the function $g(m)$ in the interval $0\leq m<1$.}\label{Energy2}
\end{figure}

The configurational part is given by
\begin{equation}
\Omega_{\mathrm{conf}}(E-K)\simeq \exp{\left( \ln \Omega_{\mathrm{conf}}\right)}.
\end{equation}
Then, the entropy is $s=\frac{1}{N}\ln \Omega$, therefore, $\Omega(E,N)$ can be expressed as
\begin{equation}
\Omega(E,N)=\frac{N}{2}\int \mbox{d}u \; \exp\!\!\left[N\left(\frac{1}{2}+\frac{1}{2}\ln(2\pi)+\frac{1}{2}\ln u+s_{\mathrm{conf}}(N \tilde{u})\right)\right],
\end{equation}
where  $\tilde{u}=U/N$ is the potential energy per particle, then $\Omega_{\mathrm{conf}}(E-K)=\Omega_{\mathrm{conf}}(N\tilde{u})$. As before, this integral in a maximal problem is given by
\begin{eqnarray}
s&\!\!=\!\!&\frac{1}{N}\ln \Omega(E,N) \\ 
\!\!&=\!\!&\frac{1}{N}\ln\!\left(\!\frac{N}{2}\!\int \!\mbox{d}u  \;\exp\!\left(\!N\!\left(\!\frac{1}{2}\!+\!\frac{1}{2}\ln(2\pi)+\frac{1}{2}\ln u+s_{\mathrm{conf}}(N \tilde{u})\!\!\right)\!\!\right)\!\!\right)\label{entr}
\end{eqnarray}
Solving the integral (\ref{entr}) in the same way as made in section \ref{evalpf},
\begin{eqnarray}
s&=&\!\!\frac{1}{N}\ln\left(\frac{N}{2}\exp{\!\left(\!\frac{N}{2}\!+\!\frac{N}{2}\ln(2\pi)\right)}\!\right)\displaystyle \!\sup_{u}\!\left(\!\frac{N}{2}\ln u\!+\!Ns_{\mathrm{conf}}(N\tilde{u}(u))\!\!\right)\\
\!\!&=&\!\!\frac{1}{N}\ln\frac{N}{2}\!+\!\frac{1}{2}+\frac{1}{2}\ln 2\pi+\displaystyle\sup_u\!\left(\!\frac{1}{2}\ln u+ s_{\mathrm{conf}}(N\tilde{u}(u))\right)
\end{eqnarray}
Further, the potential energy per particle can be expressed from Eq.(\ref{epot}), for $M_x\approx m$ and $M_y\approx 0$ as
\begin{equation}
\tilde{u}=U/N=-\lambda\left(m^2+\frac{\Delta}{2}\right),
\end{equation}
and from $U=E-K$, $\tilde{u}=\varepsilon-u/2$,  $\varepsilon=E/N$ and $u=2(\varepsilon-\tilde{u})=2(\varepsilon+\lambda(m^2+\Delta/2))$; then, in the thermodynamic limit, the entropy can be expressed
\begin{equation}
s=\frac{1}{2}+\frac{1}{2}\!\ln 2 \pi+\frac{1}{2} \ln 2+ \displaystyle\sup_m\left[\frac{1}{2}\ln (\varepsilon+\lambda(m^2+\Delta/2))+s_{\mathrm{conf}}(N\tilde{u}(u))\right].
\end{equation}
Now, we compute the configurational entropy $s_{\mathrm{conf}}$. As shown before, the term $M_y$ is negligible compared to $M_x$, this information can be introduced in $s_{\mathrm{conf}}$, as follows
\begin{equation}
\Omega_{\mathrm{conf}}=\int \displaystyle\prod_{l=1}^N \mbox{d}\theta_l \delta\left(\sum_j \cos \theta_j-Nm\right)\delta\left(\sum_j\sin\theta_j\right),
\end{equation}
it is $M_x\simeq m$, and $M_y\simeq 0$. Then we can compute expressing in the Fourier representation
\begin{equation}
\Omega_{\mathrm{conf}}\!=\!\left(\!\frac{1}{2\pi}\!\right)^{\!2}\!\!\int\!\! \mbox{d} q_1\!\int\!\!\mbox{d}q_2\!\int\! \displaystyle{\prod_{l=1}^N} \mbox{d}\theta_l {\exp}{\!\left(\!iq_1\!\sum_j \!\cos\! 
\theta_j\!-\!Nm \!\right)} \!\exp\!\!\left(\!iq_2\!\sum_j\!\sin\!\theta_j\!\right)\!\!,
\end{equation}
which corresponds to the first kind Bessel function $J_0(z)$.
\begin{equation}
\Omega_{\mathrm{conf}}=\left(\frac{1}{2\pi}\right)^2\int \mbox{d}q_1\int \mbox{d}q_2 \exp\left({N \left(
-iq_1 m+\ln(2\pi J_0(z))\right)}\right),
\end{equation}
where modulus of $z$ is $(q_1^2+q_2^2)^{1/2}$
Then, by solving the last integral in the same way as made in section \ref{evalpf}, the next equations are satisfied
\begin{eqnarray}
-im-\frac{J_1(z)}{J_0(z)}\frac{q_1}{z}&=&0,\\
-\frac{J_1(z)}{J_0(z)}\frac{q_2}{z}&=&0,
\end{eqnarray}
where the solutions are $q_2=0$, and $q_1=-i\gamma$, and $\gamma$ is the solution of equation
\begin{equation}
\frac{I_1(\gamma)}{I_0(\gamma)}=m.\label{Binv}
\end{equation}
Denoting by $B_{\mathrm{inv}}(m)$ the inverse of the Eq.(\ref{Binv}), we get in the thermodynamic limit
\begin{equation}
s_{\mathrm{conf}}=\displaystyle\lim_{N\rightarrow \infty} \frac{1}{N}\ln \Omega_{\mathrm{conf}}=-mB_{\mathrm{inv}}(m)+\ln I_0( B_{\mathrm{inv}}(m)).
\end{equation}
Using $u=2(\varepsilon+\lambda(m^2+\Delta/2))$, then
\begin{eqnarray}
s\!=\!\frac{1}{2}\!&+&\!\frac{1}{2}\!\ln 2 \pi\!+\!\frac{1}{2}\! \ln 2\! \nonumber\\
&+&\!\sup_m\! \left[\!\frac{1}{2}\!\ln(\varepsilon+\lambda(m^2+\Delta/2))\!-\!mB_{\mathrm{inv}}(m)\!+\!\ln \!I_0( B_{\mathrm{inv}}(m))\!\right]\!\!.
\end{eqnarray}
The maximal problem is given by the solution of equation
\begin{equation}
\frac{\lambda \; m}{\varepsilon+\lambda(m^2+\Delta/2)}-B_{\mathrm{inv}}(m)=0\label{gm}
\end{equation}
where we define $g(m)=\frac{m}{\varepsilon+\lambda(m^2+\Delta/2)}$, which is shown in the Figure \ref{Energy2} as the solution of the Eq.(\ref{gm}). The solution for $\varepsilon<1$ is shown in the panel ($a$). The panel ($b$) show the function $g(m)$ for $\varepsilon > 1$, where the magnetization is always $m=0$. The one solution for $m\neq0$ occurs when $\varepsilon<1$. Therefore, the Figure \ref{Energy2}  shows the function $g(m)$ for different values of $\varepsilon$.\\
\begin{figure}[ht!]
\centering
    \includegraphics[scale=0.45]{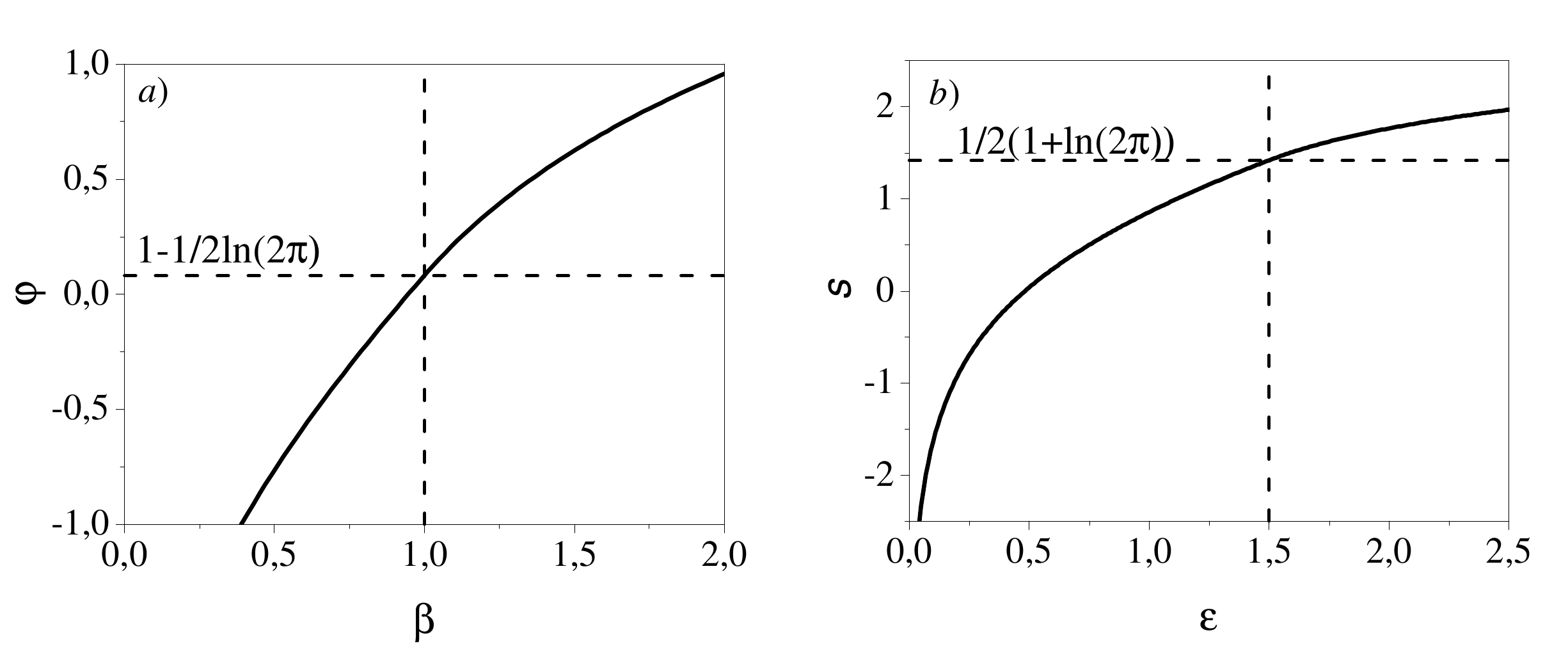}
\caption{We depict  in ({\it a})   the free energy as a function of $\beta$, in ({\it b}) the entropy $s$ as a function of the internal energy $\varepsilon$. The critical point is located in $\beta_c= 1$ and $\varepsilon_c = 3/2$ for $\Delta=-2$ and $\lambda=1$.} 
\label{Entropy}
\end{figure}

Calling $m=m(\beta)$ to the solution of extremal problem for $m$, finally the entropy can be expressed by
\begin{equation}
s\!=\!\frac{1}{2}\!+\!\frac{1}{2}\ln 2 \pi\!+\!\frac{1}{2} \ln 2\!+\! \frac{1}{2}\ln(\varepsilon\!+\!\lambda(m^2(\varepsilon)\!+\!\Delta/2))\!-\!xB_{\mathrm{inv}}(x)\!+\!\ln I_0( B_{\mathrm{inv}}(m)).
\end{equation}
The Figure \ref{Entropy} show the microcanonical entropy for the problem, and if we take the derivative respect $\varepsilon$, we recover the canonical solution for the caloric curve, i.e.,
\begin{equation}
\beta=\frac{ds}{d\varepsilon}=\frac{1}{2(\varepsilon+\lambda(m^2+\Delta/2))},\label{ccurve}
\end{equation}
which coincides with canonical solution from Eq.(\ref{ASolution}).

\section{Summary and Concluding Remarks}\label{concluding}

In physics, we find several interesting systems with analytical solutions. Some of them are related to charges, spins, rotors and dipoles as discussed in standard books of physics.
Some extra examples are introduced in recent literature \cite{ClaroyCurilef,curilef2006,AtenasAOP2014} .

In summary,  in the canonical ensemble the problem is analytically solved using Gaussian integrals as mentioned in Appendix 1, to obtain the magnetization and the inverse temperature at equilibrium as seen in Eqs.(\ref{XSolution}) and (\ref{ASolution}).  
In addition, in the microcanical ensemble, we evaluate the number of microstates to get the entropy. It is noticed that the caloric curve is reobtained from this procedure according to Eq.(\ref{ccurve}).
At this stage, it is relevant to emphasize that the solution of this system becomes analytical. However, solvable problems in statistical physics, considering long-range interactions, are not abundant. Therefore, it is pertinent to  highlight that the system of $N$ interacting dipoles in the mean field approximation is one of them. A perspective of the model is graphically shown in Figure \ref{ring} as dipoles orientated in a ring where the evolution is defined by evolving  orientation in the ring.

The phases that appear in the phase diagram are illustrated in Figure \ref{approx}(\emph{b}) and Figure \ref{Energy}, whose variables $m^2$ and $\varepsilon$
depend on $\beta$  and  $C_v$ depends on $1/\beta$.  At low energy,   a phase appears of a single cluster of particles and; at high energy, a homogeneous phase is recovered. According to Eq.(\ref{epot}) and considering $\Delta=-2$ and $\lambda=1$, we analytically obtain the parameters that define the critical point, as critical inverse temperature $\beta_c=1$, critical internal energy $\varepsilon_c=3/2$, critical magnetization $M_c=0$ as depicted in Figure \ref{approx} and Figure \ref{Energy}. 
The specific heat grows diverging with the temperature $1/\beta \rightarrow 1^-$  and keeps a constant value  $C_v = 1/2$ whether  $1/\beta > 1$, which corresponds to an ideal gas in one dimension.
The partition function is evaluated by the extremum of the Gaussian integral that is formally called the saddle point method\cite{saddle,Goutis,Huzurbazar}. 
Additionally, from the microcanonical entropy for the problem, the thermodynamic quantities are reached for recovering the canonical solution for the caloric curve.

We show that the slopes  $d\varphi/d\beta$ (Figure \ref{Entropy}(\emph{a})) and $dS/d\varepsilon$  (Figure \ref{Entropy}({\it b})), both are always positive, whenever temperature is also positive. Further, the slope in the latter figure becomes smaller as energy increases, meaning that temperature becomes bigger.  Therefore, this works out then that temperature is directly related to the energy.   
However, in systems with long range interactions,  the slope can increase with increasing energy, before the system reach the equilibrium. This fact can be relevant in dynamical studies that we will implement in the future.
As shown previously\cite{Atenas2017}, the d-HMF model can give us a good perspective for studying systems with long-range interactions. 

\section*{Acknowledgments}
We would like to thank partial financial support from FONDECYT project 1170834.
\section{Appendix}\label{app1}
Using the next identities\cite{Stratonovich1,Stratonovich2}:
\begin{eqnarray}
\sqrt{\frac{\pi}{b}}&=&\int^{\infty}_{-\infty} \mathrm{d}x\: \mathrm{exp}\left(-b(x-M_x)^2\right)\\
\mathrm{exp}\left(b M^2_x\right)&=&\sqrt{\frac{b}{\pi}}\int^{\infty}_{-\infty} \mathrm{d}x\: \mathrm{exp}\left(-bx^2+2bM_xx\right)\label{emx}
\end{eqnarray}
and
\begin{eqnarray}
\sqrt{2 \pi b}&=&\int^{\infty}_{-\infty} \mathrm{d}y\: \mathrm{exp}\left(-\frac{1}{2}(\frac{y}{\sqrt{b}}+i\sqrt{b}{M_y})^2\right)\\
\mathrm{exp}\left(-\frac{b M^2_y}{2}\right)&=&\sqrt{\frac{1}{2\pi b}}\int^{\infty}_{-\infty} \mathrm{d}y \: \mathrm{exp}\left(-\frac{y^2}{2b}-iM_y y\right)\label{emy}
\end{eqnarray}
Applying to the current problem $b=\beta \lambda N$. In equilibrium, it is expected a symmetric distribution of orientations $\rho(\theta)$; therefore, $M_y=\frac{\sum_i \sin\theta_i}{N} \approx \int_{0}^{2\pi} \mbox{d}\theta \sin \theta \rho(\theta)$ vanishes  when $N$ is large, then we have:
\begin{equation}
\exp\left(\beta \lambda N M_x^2\right)=\sqrt{\frac{\beta \lambda N }{\pi}}\int^{\infty}_{-\infty} \mathrm{d}x \exp{( -\beta \lambda N x^2+2\beta  \lambda  x \sum_i\cos\theta_i)}\label{emxx}
\end{equation}
and 
\begin{equation}
 \exp\left(-\beta N\frac{\lambda}{2}M_y^2\right)= \frac{1}{\sqrt{2 \pi \beta \lambda N }}\int^{\infty}_{-\infty} \mathrm{d}y \exp{( -\frac{ y^2}{2\beta \lambda N})}=1\label{emyy}
\end{equation}


\vspace{1cm}
\section*{References}
\begin{thebibliography}{88}
\bibitem{reichl} L. Reichl, ``\emph{A Modern Course in Statistical Physics}", $4^\mathrm{th}$ Edition (Wiley 2016)
\bibitem{soto} R. Soto, ``\emph{Kinetic Theory and Transport Phenomena}", $1^\mathrm{st}$ Edition (Oxford Master Series in Physics, 2016)
\bibitem{Glasser} M. L. Glasser, ``\emph{Exact Partition Function for the Two-Dimensional Ising Model}", Am. J. Phys. 38, 1033 (1970)
\bibitem{DS} D. Stauffer, ``\emph{Social applications of two-dimensional Ising models}", Am. J. Phys. 76, 470 (2008)
\bibitem{Tatekawa} T. Tatekawa, F. Bouchet, T. Dauxois and S. Ruffo, ``\emph{Thermodynamics of the self-gravitating ring model}", Phys. Rev. {\bf E 71}, 056111 (2005)
\bibitem{cdr} A. Campa, T. Dauxois, S. Ruffo, ``\emph{Statistical mechanics and dynamics of solvable models with long-range interactions}",
Physics Reports {\bf 480}, 57-159 (2009)
\bibitem{Pluchino} A. Pluchino, V. Latora and A. Rapisarda, ``\emph{Dynamics and thermodynamics of a model with long-range interactions}",
Continuum Mech. Thermodyn. {\bf 16}, 245-255 (2004).
\bibitem{Curilef} S. Curilef, ``\emph{A long-range ferromagnetic spin model with periodic boundary conditions}", Phys. Lett. {\bf A 299}, 366-370 (2002)
\bibitem{TDauxois} T. Dauxois, S. Ruffo, E. Arimondo, M. Wilkens (Eds.), ``\emph{Dynamics and Thermodynamics in Systems with Long-range Interactions}, Lecture Notes in Physics, Vol. 602, Springer, Berlin, 2002.
\bibitem{Levin}   Y. Levin, R. Pakter, F.B. Rizzato, T.N. Teles and F.P. da C. Benetti, ``\emph{Nonequilibrium statistical mechanics of systems with long-range interactions}'',
Physics Reports {\bf 535}, 160 (2014)
\bibitem{delPinoPRB2007}  L.A. del Pino, P. Troncoso and S. Curilef, ``\emph{Thermodynamics from a scaling Hamiltonian}, Phys. Rev. {\bf B 76}, 172402 (2007).
\bibitem{kac} M. Kac, G. Uhlenbeck, P.C. Hemmer, `\emph{On the van der Waals Theory of the Vapor-Liquid Equilibrium. I. Discussion of a One-Dimensional Model}", J. Math. Phys. {\bf 4}, 216 (1963)
\bibitem{Atenas2017} B. Atenas and S. Curilef, ``\emph{Dynamics and thermodynamics of systems with long-range dipole-type interactions}", Phys. Rev. {\bf E 95}, 022110 (2017)
\bibitem{stanley}  H. E. Stanley, ``\emph{Introduction to phase transition and critical phenomena}", Oxford University Press 1971.
\bibitem{ClaroyCurilef} S. Curilef and F. Claro, ``\emph{Dynamics of two interacting particles in a magnetic field in two dimensions}", Am. J. Phys. {\bf 65}, 244-250 (1997)
\bibitem{curilef2006} P. Troncoso and S. Curilef, ``\emph{Bound and trapped states of an electric dipole in a magnetic field}," Eur. J. Phys. {\bf 27}, 1315-1322  (2006).
\bibitem{AtenasAOP2014} B. Atenas, L. A. del Pino and S. Curilef, ``\emph{Classical states of an electric dipole in an external magnetic field: Complete solution for the center of mass and trapped states}", Annals of Physics {\bf 350}, 605-614 (2014)
\bibitem{saddle} M. Le Bellac, F. Mortessagne and G. G. Batrouni,  ``\emph{Equilibrium and nonequilibrium statistical thermodynamics}", $1^\mathrm{st}$ Edition (Cambridge University Press 2004)
\bibitem{Goutis} C. Goutis and G. Casella, ``\emph{Explaining the Saddle point Approximation}", The American Statistician {\bf 53}, 216-224 (1999)
\bibitem{Huzurbazar} S. Huzurbazar, ``\emph{Practical Saddlepoint Approximations}", The American Statistician  {\bf 53}, 225-232 (1999)
\bibitem{Stratonovich1} R.L. Stratonovich, ``\emph{On a method of calculating quantum distribution functions}", Soviet Physics Doklady {\bf 2}, 461 (1958)
\bibitem{Stratonovich2} J. Hubbard, ``\emph{Calculation of partition functions}", Phys. Rev. Lett. {\bf 3}, 77„-78 (1959)
\end{thebibliography}
\end{document}